\documentclass[%
reprint,
superscriptaddress,
aps,
prl,
portrait,
]{revtex4-2}
\usepackage{blindtext}
\usepackage{amsmath}
\usepackage{bm}
\usepackage{xcolor}
\usepackage{soul}
\usepackage{siunitx}
\usepackage[version=4]{mhchem} 
\usepackage{graphicx}
\usepackage{subfig}
\usepackage[normalem]{ulem}
\usepackage{pdfpages}
\graphicspath{{./figures/}}

\makeatletter
\AtBeginDocument{\let\LS@rot\@undefined}
\makeatother

\begin{document}

\preprint{APS/123-QED}

\title{Surface charge deposition by moving drops reduces contact angles}

\author{\underline{Xiaomei Li}}
\thanks{X.L. and A.D.R. contributed equally to this work.}
\affiliation{
 Max Planck Institute for Polymer Research, Ackermannweg 10, 55128 Mainz, Germany 
}

\author{\underline{Aaron D. Ratschow}}
\thanks{X.L. and A.D.R. contributed equally to this work.}
\affiliation{
 Institute for Nano- and Microfluidics, TU Darmstadt,\\
 Alarich-Weiss-Stra{\ss}e 10, D-64237 Darmstadt, Germany
}

\author{Steffen Hardt}
\email{hardt@nmf.tu-darmstadt.de}
\affiliation{
 Institute for Nano- and Microfluidics, TU Darmstadt,\\
 Alarich-Weiss-Stra{\ss}e 10, D-64237 Darmstadt, Germany
}

\author{Hans-Jürgen Butt}
\email{butt@mpip-mainz.mpg.de}
\affiliation{
 Max Planck Institute for Polymer Research, Ackermannweg 10, 55128 Mainz, Germany 
}

\date{\today}

\begin{abstract}
Slide electrification - the spontaneous charge separation by sliding water drops - can lead to an electrostatic potential of \SI{1}{kV} and change drop motion substantially. To find out, how slide electrification influences the contact angles of moving drops, we analyzed the dynamic contact angles of aqueous drops sliding down tilted plates with insulated surfaces, grounded surfaces, and while grounding the drop. The observed decrease in dynamic contact angles at different salt concentrations is attributed to two effects: An electrocapillary reduction of contact angles caused by drop charging and a change in the free surface energy of the solid due to surface charging. 
\end{abstract}

\maketitle

The movement of liquid drops on solid surfaces plays a fundamental role in many natural and technological processes. Examples range from the spreading of raindrops on plant leaves or glass to processes like inkjet printing or coating \cite{gennes2004capillarity,bonn2009wetting,lohse2022fundamental}. The interaction between a liquid and a solid is largely determined by the contact angle near the three-phase contact line. Young’s equation relates the contact angle ($\theta$) to the interfacial energies of the liquid surface (L), solid surface (S), and the solid-liquid interface (SL) with \cite{young1805iii},
\begin{equation}
    \gamma_\mathrm{L} \cos(\theta)=\gamma_\mathrm{S}-\gamma_\mathrm{SL}.
\end{equation}
A lower contact angle indicates a higher solid surface energy. Therefore, surface wettability can be controlled by choosing a high or low surface energy material, which leads to low or high contact angles, respectively. The composition of a smooth surface determines its contact angle. To control contact angles, electrowetting is a versatile tool. It is used in various microfluidic applications \cite{mugele2018electrowetting}. In electrowetting, the contact angle of a sessile drop on a dielectric substrate on top of an electrode decreases when a voltage ($\Delta U$) is applied between the drop and the electrode. Microscopically, this effect is due to the electrostatic Maxwell stress acting on the liquid surface in the close vicinity of the contact line. Macroscopically, the effect can be attributed to a change in effective free surface energy of the solid-liquid interface because of the accumulation of charges \cite{buehrle2003interface,mugele2007equilibrium}. Intuitively, it is energetically favorable for counter-charges to accumulate at the solid-liquid interface under an applied potential, and thus its surface energy is reduced compared to the case without charges. The change in the solid-liquid interfacial energy can be expressed as
\begin{equation}
    \Delta \gamma_\mathrm{SL}=\gamma_\mathrm{SL}^\mathrm{eff}-\gamma_\mathrm{SL}=-\frac{\varepsilon_0\varepsilon_r}{2d}\Delta U^2.
\end{equation}
It depends on the permittivity ($\varepsilon_0=$ vacuum permittivity, $\varepsilon_r=$ relative permittivity) and the thickness of the substrate $d$ \cite{mugele2018electrowetting} separating the liquid from an electrode. In the macroscopic description, the change in contact angle is given by the Young-Lippmann equation \cite{buehrle2003interface},
\begin{equation}
    \cos(\theta)-\cos(\theta')=\frac{\Delta \gamma_\mathrm{SL}}{\gamma_\mathrm{L}}.
\end{equation}
Here $\theta$ and $\theta'$ are the contact angles without and with an applied voltage.  

Another physical phenomenon that involves drops and electrostatic charges is slide electrification \cite{yatsuzuka1994electrification,stetten2019slide,li2022spontaneous,Jin.2022Electrification}. A sliding aqueous drop on a hydrophobic surface can acquire a net charge while leaving behind an opposite charge on the dewetted surface. On low permittivity, hydrophobic surfaces, the drop is usually positively charged and negative surface charges are left behind \cite{beattie2006intrinsic,kudin2008water,tian2009structure,preovcanin2012surface}. Spontaneous charging of moving drops influences their motion substantially by direct Coulomb forces between the charges in the drop and the opposite charges on the solid surface \cite{li2022spontaneous,diaz2023self}. However, it is still not clear if spontaneous charging changes the contact angle. Here, we address the question: Do charges in the drop and/or surface charges generated by slide electrification change the advancing and receding contact angles? If yes, how does this effect depend on the salt concentration?
\\
To answer these questions, we imaged sliding aqueous drops (Supporting information, S1) in a custom-made tilted plate setup (Fig. 1a) \cite{li2022spontaneous,li2023drop}. Drops with a volume of \SI{30}{\micro L} were placed onto a tilted surface by a peristaltic pump with a grounded syringe needle at fixed intervals of \SI{1.5}{s}. The surfaces used were flat, smooth, and hydrophobic, with an average roughness $<\SI{1}{nm}$ within an area of $0.5\times\SI{0.5}{\micro m^2}$ (Supporting information: S2 and S3). Every drop moving down the surface first contacted a grounded electrode. We imaged the sliding drops from the side with a high-speed camera and set the slide length and time to zero when drops detach from the grounded electrode and enter the recording window. At this point, they already have an initial velocity. Based on the side-view images, the positions, velocities, and contact angles of the advancing and receding contact lines were determined automatically by an adapted image analysis MATLAB code \cite{andersen2017drop}. The drop velocity for every drop position was defined as the mean of the velocities of the front and rear contact lines \cite{li2021adaptation}. As surfaces, we prepared \SI{60}{nm} thick Teflon films on quartz plates (Teflon-quartz) by dip-coating (\SI{1}{cm/min}) from a solution of \SI{1}{wt\%} Teflon AF 1600 and annealing at 160 °C under vacuum for 24 h. The quartz plates were \SI{1}{mm} thick and placed on a grounded metal plate. 
\\

\begin{figure}
	\includegraphics[width=0.99\columnwidth]{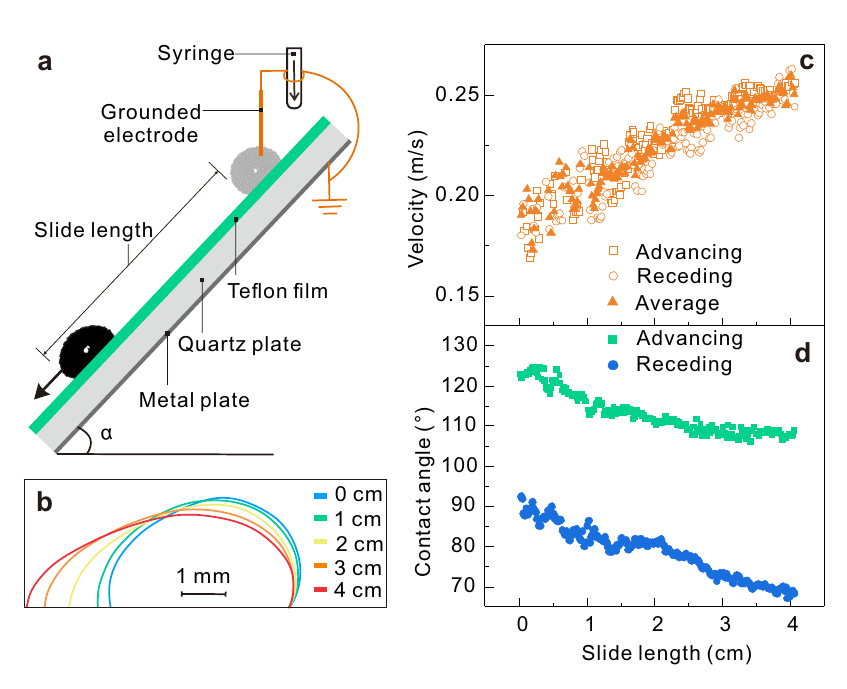}
  \caption{Experiment for aqueous drops with \SI{1}{mM} \ce{NaCl} on a 40° tilted initially uncharged Teflon-quartz surface. The drop was grounded until it detached from the needle and  ungrounded during the whole sliding. (a) Schematic of the experimental setup. (b) Drop profiles for different values of the slide length, (c) drop velocity, (d) dynamic advancing and receding contact angle over the slide length of the drop shown in (a).}
\end{figure}

When placing an aqueous drop containing \SI{1}{mM} \ce{NaCl} on a pristine, uncharged Teflon-quartz surface (Fig. 1a), the drop accelerates. Its shape becomes more elongated (Fig. 1b) while the velocity increases (Fig. 1c). In addition, the dynamic advancing and receding contact angles decrease with increasing velocity (Fig. 1d). Established theories, such as the Cox-Voinov hydrodynamic model \cite{voinov1976hydrodynamics,cox1986dynamics}, the molecular kinetic model \cite{blake1969kinetics},  combinations of both \cite{brochard1992dynamics,petrov1992combined}, and the adaptation model \cite{Butt.2018} predict a decrease in receding, but an increase in advancing contact angle with increasing velocity. This prediction does not agree with our measurements. We conclude that there are additional effects influencing the contact angles and propose that the change in contact angle is due to charging of the drops. 

To verify that charging of aqueous drops causes this change in contact angles, we sputter-coated the quartz plates with \SI{5}{nm} chromium and \SI{35}{nm} gold before coating Teflon films on top (Teflon-gold). In earlier experiments we had shown that in contrast to Teflon-quartz (Fig. 2a), charging effects are negligible for \SI{50}{nm} polymer films on grounded gold (Fig. 2b) \cite{li2022spontaneous, diaz2023self}. For Teflon-gold, the advancing contact angle indeed increase with velocity (Fig. 2e, orange symbols) and the decrease of the receding contact angle is weaker (Fig. 2f, orange symbols). 

We propose that electrowetting reduces the contact angles of charged drops. The sliding drop on the Teflon-quartz surface spontaneously acquires positive charges, leaving negative surface charges behind. The related electrostatic potential leads to an electrowetting effect. To support this hypothesis, we calculate $\Delta \gamma_\mathrm{SL}$ and test, if the anticipated changes in contact angle (equation 3) are large enough. First, we convert measured drop charges $Q$ to a potential $\Delta U=Q/C$ with the capacitance of the drop $C=(A\varepsilon_0\varepsilon_r)/d$ (Supporting information, S4). Here, $A$ is the contact area of the drop. In previous measurements \cite{li2022spontaneous}, after \SI{4}{cm} sliding on a Teflon-quartz surface the drop charge was $Q\approx\SI{0.7}{nC}$, $A\approx\SI{17}{mm^2}$ and $\varepsilon_r=4.5$, we estimate $\Delta U\approx\SI{1}{kV}$, comparable to potentials reported by \cite{Xu.2022Triboelectric}. Based on equation 2, the changes in the solid-liquid interfacial tension are of the order of \SI{10}{mN/m} leading to a decrease of $\approx9$° in contact angles. Since the potential continuously increases with increasing slide length, electrowetting can explain the decrease in advancing contact angle. 
 
Are there other electrostatic effects influencing the contact angles? To isolate such effects, we use the same Teflon-quartz surfaces as previously but constantly ground the sliding drop with a tungsten wire to prevent drop charging and electrowetting effects (Fig. 2c). The grounded tungsten wire (\SI{25}{\micro m} diameter) was spanned parallel to the surface at $\approx\SI{1}{mm}$ height along the path of the drop. Its influence on the drop velocity or contact angles was negligible (Supporting information, S5). With the grounded wire, the drop can still deposit negative surface charge at its rear, but the drop itself remains uncharged. 

Fig. 2d-f shows the velocity, the dynamic advancing, and the dynamic receding contact angles versus slide length for the $1^{st}$ and $100^{th}$ consecutive grounded drop on the Teflon-quartz surface. We observe a distinct difference between the $1^{st}$ (green circles) and $100^{th}$ (blue triangles) drop. This observation indicates that the surface charge on the solid-air interface influences the contact angles and drop motion, even if the drop is uncharged. For comparison, the results of the reference measurement on the Teflon-gold surface (orange circles) are also plotted. There is no significant difference between the $1^{st}$ and $100^{th}$ drop on the gold substrate (Supporting information, S6). 
\\

\begin{figure}
	\includegraphics[width=0.99\columnwidth]{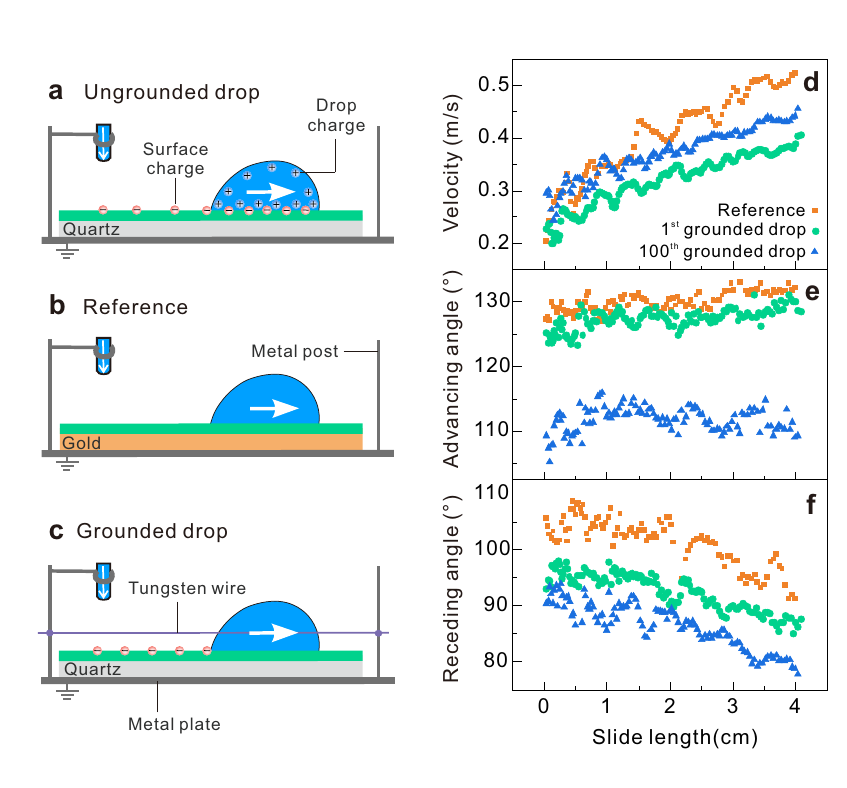}
  \caption{Dynamic contact angles reduced by slide electrification. Schematics of a drop (\SI{30}{\micro L}  drop with \SI{1}{mM} \ce{NaCl}) sliding down the 40° tilted (a) A Teflon-quartz surface and (b) a Teflon-gold surface without drop grounding (b), and (c) a Teflon-quartz surface with  drop grounding during sliding. The corresponding drop velocity (d), dynamic advancing contact angle (e), and dynamic receding contact angle (f) over slide length for the $1^{st}$ and $100^{th}$ consecutive drop. }
\end{figure}

Compared with the Teflon-gold reference (Fig. 2e-f, orange), the dynamic advancing contact angle of the $1^{st}$ grounded drop on the Teflon-quartz surface was not affected. Only the dynamic receding contact angle was reduced by 10°. In comparison, the dynamic advancing contact angle of the ungrounded drop (Fig. 1d) decreases for slide lengths $>0$ as the ungrounded drop charged and electrowetting commenced. For the $100^{th}$ grounded drop on Teflon-quartz (Fig. 2e-f, blue), both the dynamic advancing and receding contact angles deviate from the Teflon-gold reference. The main difference was that the grounded drop continuously deposited charges at its receding contact line (Fig. 2b-c) while the drop on the Teflon-gold surface did not generate surface charges. Thus, in addition to electrowetting, surface charges cause a fundamentally new electrostatic effect that decreases contact angles.

We propose that charges on the solid-gas interface increase the surface energy and according to Young’s equation reduce the contact angles. The surface energy is increased by two effects. The first is the self-energy of the charges on the surface, also referred to as Born energy. The corresponding change in the surface energy is of the order of \SI{10}{\micro N/m} (Supporting information, S7) and is thus negligible. The second effect is that charges on the surface repel each other by Coulomb interaction. Thus, forming a layer of charges requires electrostatic work. To derive a theoretical scaling for this effect, we analytically calculate the work required to deposit an additional elementary charge on an already-charged surface. This energy depends on the size of the charged patch. As an example, we consider a circular patch of charges of radius $R$ and a charge density $\sigma$. After area-averaging this energy, we obtain the change in free surface energy of the solid due to the presence of a charge density  $\sigma$ (Supporting information, S7):
\begin{equation}
    \Delta\gamma_\mathrm{S}=\gamma_\mathrm{S}^\mathrm{eff}-\gamma_\mathrm{S}=\frac{\sigma^2R}{\varepsilon_0(1+\varepsilon_r)}.
\end{equation}
The corresponding change in contact angle is given by
\begin{equation}
    \cos(\theta)-\cos(\theta')=-\frac{\Delta\gamma_\mathrm{S}}{\gamma_\mathrm{L}}.
\end{equation}
The surface energy increases quadratically with the charge and linearly with the length scale of the charged area $R$. For the $1^\mathrm{st}$ drop, there is only one characteristic scale of the problem that comes into consideration for $R$, which is the drop size, represented by its radius. With $R=\frac{2}{mm}$ and a charge density of $\sigma=\SI{10}{\micro C/m^2}$ \cite{li2022spontaneous}, we estimate an increase in solid surface energy of around \SI{10}{mN/m}, which would substantially change contact angles. 

A macroscopic description with Young’s equation and the effective solid surface energy $\gamma_\mathrm{S}^\mathrm{eff}$ is only viable above the characteristic length scale of the microscopic effects. On the microscopic scale, electrostatic forces, expressed by the Maxwell stress, and capillary forces balance at the liquid-gas interface. Mathematically, the electrostatic problem of an isopotential wedge, representing the liquid, next to a charged surface does not have an inherent length scale. Consequently, there is no apparent length scale over which the Maxwell stress is localized. It even becomes singular at the contact line \cite{li2022spontaneous,yeo2005static} for the model problem of an isopotential wedge. However, such mathematical singularities do not occur in nature. There are different mechanisms that could introduce a microscopic length scale close to the contact line. First, we have to consider that the treatment of the liquid surface as isopotential only applies on length scales above the Debye length, with $\lambda\approx1-\SI{100}{nm}$ in aqueous solutions. Moreover, singularities of the electric field at the contact line would lead to electrostatic discharge \cite{vallet1999limiting,lowe2020impact} above the limiting field strength of humid air, which is $\approx\SI{2}{MV/m}$ \cite{li2018effect}. Following this argument, the singularity is eliminated on the length scale where electrostatic discharge first occurs. With numerical simulations, we show that the introduction of such a microscopic length scale strongly localizes the Maxwell stress, which makes a macroscopic description with Young’s equation viable. We estimate the limiting length scale for the macroscopic description of the effect to be of the order of \SI{1}{\micro m} (Supporting information, S8), above which the contact angle should be well-defined by a change in the effective solid surface energy, as shown in Fig. 3.
\\
 
\begin{figure}
	\includegraphics[width=0.99\columnwidth]{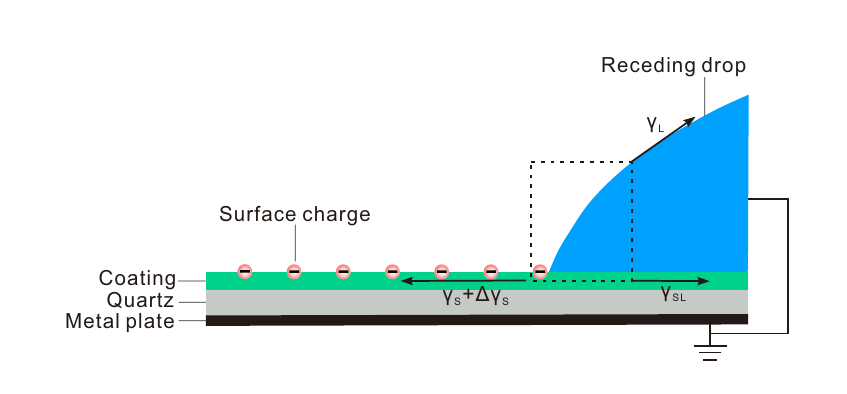}
  \caption{Schematic representation of the discovered effect of contact-angle modification via electrostatic interaction with surface charges. The sliding drop deposits charges on the solid surface, which deform the liquid surface via electrostatic interactions. Above a characteristic length scale ('control' volume indicated by a dashed rectangle), the effect can be subsumed as a change in solid surface energy.}
\end{figure}

To compare theory and experiments and explain the change of dynamic contact angles, we consider three effects: (i) Non-electrostatic contributions such as hydrodynamics described by the Cox-Voinov model, contact-line friction because of local pining and de-pining of contact line, and adaptation, (ii) surface charge-induced changes of the solid surface (equation 4), and (iii) electrowetting due to charging of the drop (equation 2). Our experiments are designed in such a way that the reference measurement on Teflon-gold substrates is only influenced by (i). On Teflon-quartz substrates the grounded drop is influenced by (i) and (ii), and the ungrounded drop is influenced by (i)-(iii). The initial decrease in receding contact angle between the reference and the grounded drop for the $1^{st}$ drop was around 10° (Fig. 2f). To fully explain this with equation 4, the drop with a radius of \SI{2}{mm} on quartz ($\varepsilon_r=4.5$) would have to deposit a surface charge of $\sigma=\SI{16}{\micro C/m^2}$. This value agrees magnitude-wise with our previously published measurement of \SI{10.3}{\micro C/m^2} on the same substrate for the first drop \cite{li2022spontaneous}. 

For the $100^{th}$ drop (blue) shown in Fig. 2e-f, also the advancing contact angle decreased. We attribute this to surface charges left behind by previous drops. Due to hydrodynamics, contact-line friction, and adaptation (effect i), receding contact angles are lower than advancing contact angles. We observe that dynamic receding contact angles are more affected by surface charge than advancing ones. In line with this observation, a calculation of the contact angle change as a function of surface charge density for different initial contact angles shows that lower contact angles are more affected (Fig. 4a). Note that the applicability of such models becomes questionable for contact angles of 20-30° due to electrostatic discharge \cite{vallet1999limiting}.  
 
To demonstrate the universality of the effect, we measured surfaces with different coatings and drops with different salts. For comparison, we calculated $\cos(\theta')-\cos(\theta)$. We observe it on \SI{35}{nm} thick polystyrene (PS) films coated quartz plates, molecular layers of perfluoroctyltriethoxysilane (PFOTS), and polydimethylsiloxane (PDMS) grafted to quartz plates. The effect also occurs for all salts tested (Fig. 4b and Fig. S10a). The reduction of dynamic contact angles increase with increasing salt concentration up to $\approx\SI{1}{mM}$ followed by a decrease (Fig. 4c and Fig. S10b). This trend is consistent with the reported trends of drop/surface charges \cite{helseth2020influence,sosa2020liquid,helseth2023ion}. The initial increase of the effect with salt concentration can be explained by the Péclet number dependency of charge separation. In  the drop there is a flow component directed upward at the receding contact line. It drives counterions away from the surface and extends the effective screening length. Assuming charge regulation at the solid-liquid interface, an extended screening length reduces the surface charge directly at the receding contact line which also reduces the surface charge transferred to the free solid surface. This effect is only effective if convective transport is stronger than diffusion of ions. The Péclet number $Pe=v\lambda/D$ ($v=$ drop velocity, $D=$ ion diffusivity) measures convective transport, which is more or less negligible up to $Pe=1$ and causes a decrease of charge separation for $Pe>1$. \cite{ratschow2023how} For typical values $v=\SI{0.3}{m/s}$ and $D=\SI{2e-9}{m^2/s}$, a transition between the two regimes is found at a salt concentration of \SI{2}{mM}, which explains the reduced contact angle changes at \SI{1}{mM} and below. The experimental trends are in accordance with the theoretical scaling.

The theory also predicts a scaling of the effect with the length scale associated with the charged area, that for the first drop corresponds to the drop radius $R$. To confirm this scaling, we measured grounded drops of different volumes $V$ and observe a clear increase of the effect for larger drops (Fig. 4d). We apply the theoretical scaling from the data point at \SI{10}{\micro L} on, which corresponds to a relationship $R\propto V^{(1/3)}$, and find an agreement up to $V\approx\SI{30}{\micro L}$. For larger volumes, the simple scaling breaks down as the drop height approaches the capillary length and the radius increases beyond the value it takes without the influence of gravity, which is reflected in the higher experimental values.
\\
 
\begin{figure}
	\includegraphics[width=0.99\columnwidth]{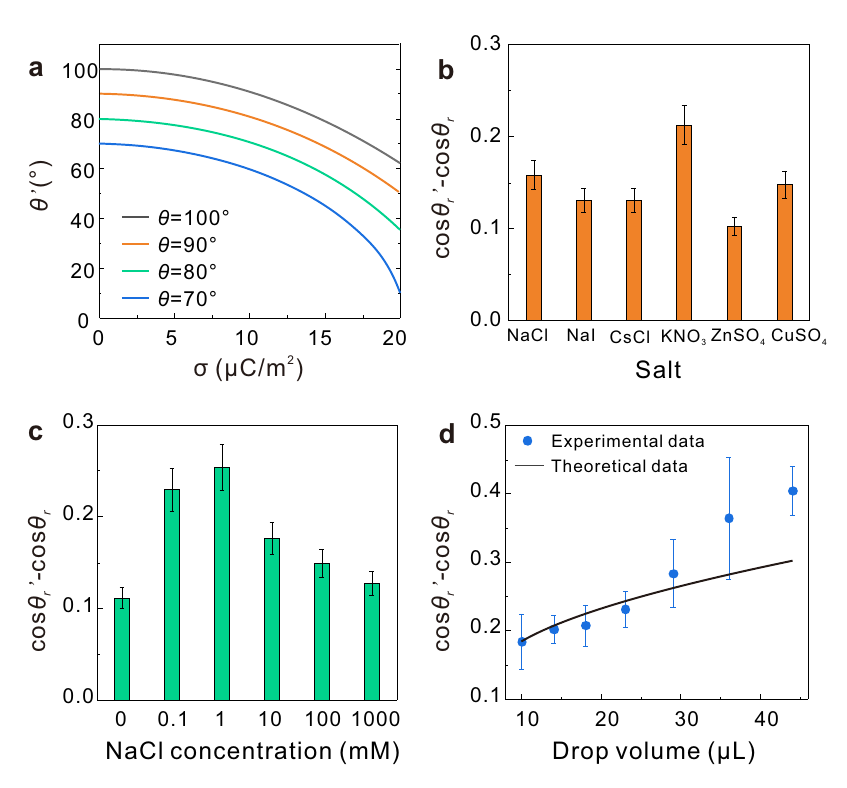}
  \caption{Universality of the effect. (a) Theoretically expected contact angle as a function of surface charge density for different initial contact angles, calculated with equations (4) and (5). Experimentally determined $\cos(\theta_r ')-\cos(\theta_r)$ for the first grounded drops with (b) different salt types, (c) salt concentrations, and (d) drop volumes of \SI{1}{mM} \ce{NaCl} aqueous solution on pristine Teflon-quartz surfaces in the velocity range $0.3-\SI{0.4}{m/s}$. $\theta_r$ and $\theta_r '$ are the dynamic receding contact angles without and with the influence of slide electrification. }
\end{figure}

In addition to our own measurements, our theory helps to explain observations from the literature. For instance, Mugele et al. \cite{mugele2005electrowetting} reported that the contact angle of an aqueous solution on Teflon permanently decreased by 5-10° after the first wetting-dewetting cycle. This was likely caused by charges deposited onto the previously uncharged surface during the initial dewetting. Sun et al. \cite{sun2019surface} experimentally demonstrated that drops move along surface charge gradients towards higher charged regions and even do so against gravity. This phenomenon is easily conceivable with equation 4, as the higher charged regions have an increased free surface energy.

To conclude, we identified two mechanisms explaining how slide electrification can lead to a reduction in dynamic contact angles. Charges in the drop induce an electric field between the drop and the subsurface electrode, which via electrowetting causes a reduction of the advancing and receding contact angles. Charges on the solid surface effectively increase the surface energy and thus reduce the contact angle according to Young’s equation. Depending on the distribution of surface charges the advancing or receding side can be affected. The latter effect can substantially reduce the dynamic contact angle, even when the drop itself is prevented from charging. We propose an analytical model based on Young’s equation, which agrees well with our experimental data. The universality of the effect is supported by measurements with different salt types, salt concentrations, drop volume, and hydrophobic coatings. The discovered effect could help to explain contact angle hysteresis in many practical cases and facilitate the design of functional surfaces by focusing on the prevention of charge separation.

\begin{acknowledgments}
We wish to thank R\"udiger Berger, Diego Diaz, and Lisa S. Bauer for their valuable suggestions regarding the experiments and Tobias Baier for helpful discussions on the theory.
H.-J.B. proposed and supervised the work, X.L. designed, conducted, evaluated, and interpreted the experiments, A.D.R. developed the theoretical framework with input from H.-J.B. and S.H. and derived the analytical model, A.D.R. and S.H. worked out the simulation model and conducted the simulations, X.L. and A.D.R. prepared the manuscript. This work was supported by the European Research Council (ERC) under the European Union’s Horizon 2020 research and innovation program (grant agreement no. 883631) (H.-J. Butt), the Priority Programme 2171 ‘Dynamic wetting of flexible, adaptive and switchable surfaces’ (grant no. BU 1556/36: X. Li, H.-J. Butt), the Department for Process and Plant Safety of Bayer AG, Leverkusen, Germany (A. D. Ratschow), and the German Research Foundation (DFG) within the Collaborative Research Centre 1194 “Interaction of Transport and Wetting Processes”, Project- ID 265191195, subproject A02b (S. Hardt) and C07 (H.-J. Butt). 
\end{acknowledgments}

\bibliographystyle{apsrev4-1} 
\bibliography{main.bbl}

\clearpage


\section*{Supplementary material}

# Surface charge deposition by moving drops reduces contact angles
## -Supplemental Material-

Xiaomei Li,[1, *] Aaron D. Ratschow,[2, *] Steffen Hardt,[2, †] and Hans-Jürgen Butt[1, ‡]

[1] *Max Planck Institute for Polymer Research, Ackermannweg 10, 55128 Mainz, Germany*
[2] *Institute for Nano- and Microfluidics, TU Darmstadt,*
*Alarich-Weiss-Straße 10, D-64237 Darmstadt, Germany*
(Dated: April 14, 2023)

**Contents**

## S1:  Liquid preparation.

The chemicals used to prepare salt solutions include distilled water ($< 1\,\mu S/cm$; Gibco, Thermo Fisher Scientific), $1\,M$ NaCl aqueous solution (Carl Roth, Germany), $1\,M$ KNO$_3$ aqueous solution (Carl Roth, Germany), $0.1\,M$ ZnSO$_4$ aqueous solution (Fluka, Germany), $1\,M$ NaOH aqueous solution (VWR International, France), 37% HCl aqueous solution (Sigma-Aldrich), NaI (99.999%, Sigma-Aldrich), CsCl (99.9%, Sigma-Aldrich), CuSO$_4$ (99%, Sigma-Aldrich). Salt solutions were prepared by mixing appropriate amounts with distilled water.

## S2:  Surface preparation.

*Substrates cleaning.*  Gold, glass slides ($76.2 \times 25.4 \times 1.0\,mm^3$, Paul Marienfeld), Si wafers ($< 0.005\,\Omega cm$; thickness, $525 \pm 25\,\mu m$, P++<100>, Silicon Materials) and quartz slides ($76.2 \times 25.4 \times 1.0\,mm^3$, Thermo Fisher Scientific) were used as substrates. Before use, they were cleaned in an ultrasonic bath in toluene and ethanol for 10 min each. After drying by N$_2$ blowing, they were O$_2$-plasma cleaned at 300 W for 10 min (Femto low-pressure plasma system, Diener electronic). Gold substrates with 5 nm chromium and 35 nm gold on glass slides were prepared by sputter coating and used immediately without further cleaning.

*Surface preparation.*  (1) 60 nm Teflon coatings on gold and quartz substrates were prepared by dip-coating with a pulling speed of 10 mm/min from a solution of 1(wt)% Teflon AF 1600 ($\varepsilon_r = 1.9$; Sigma-Aldrich) in FC-43 (Sigma-Aldrich). Before use, we annealed the Teflon samples in the oven at 160 under vacuum for 24 h. (2) 35 nm PS coatings on gold and quartz substrates were prepared by dip-coating with a pulling speed of 90 mm/min from a solution of 1(wt)% PS (molecular weight, 192 kg/mol, $\varepsilon_r = 2.6$; Sigma-Aldrich) in toluene (99.8%, Sigma-Aldrich). Before use, we annealed the PS samples at 120 °C under vacuum for 24 h. (3) PFOTS coatings on Si wafers and quartz substrates were prepared by chemical vapor deposition. Cleaned Si wafers and quartz slides were put into a vacuum desiccator containing a tiny glass bottle with 0.5 ml 1H, 1H, 2H, 2H-perfluorooctadecyltrichlorosilane (97%, Sigma-Aldrich). The desiccator was evacuated to $< 100$ mbar. After 30 min, the samples were removed and cleaned by rinsing with ethanol to remove any unbound silanes. (4) PDMS layered costings were prepared by the "grafting to" method using silicone oil (molecular weight, 6 kg/mol; Alfa Aesar). A few drops of silicone oil were deposited on the cleaned Si wafer and quartz slides. The samples were stored at $22 - 23$ °C and 30–60% relative humidity for 24–48 hours after the PDMS drops spread and covered the substrates. Before use, they were cleaned using ultrasound in toluene, ethanol, and distilled water for 10 min each to remove the unbound silicone oil.

## S3:  SFM imaging.

We used tapping mode (Dimension Icon, Bruker) to measure the morphology of all surfaces within an area of $1 \times 1\,\mu m^2$ (Fig. S1). The cantilever had a nominal resonance frequency of 300 kHz and a spring constant of 26 N/m (160AC-NA, OPUS). The errors of root-mean-squared (RMS) roughness are the standard deviation of the RMS roughness from three measurements on different positions and different patches of samples.

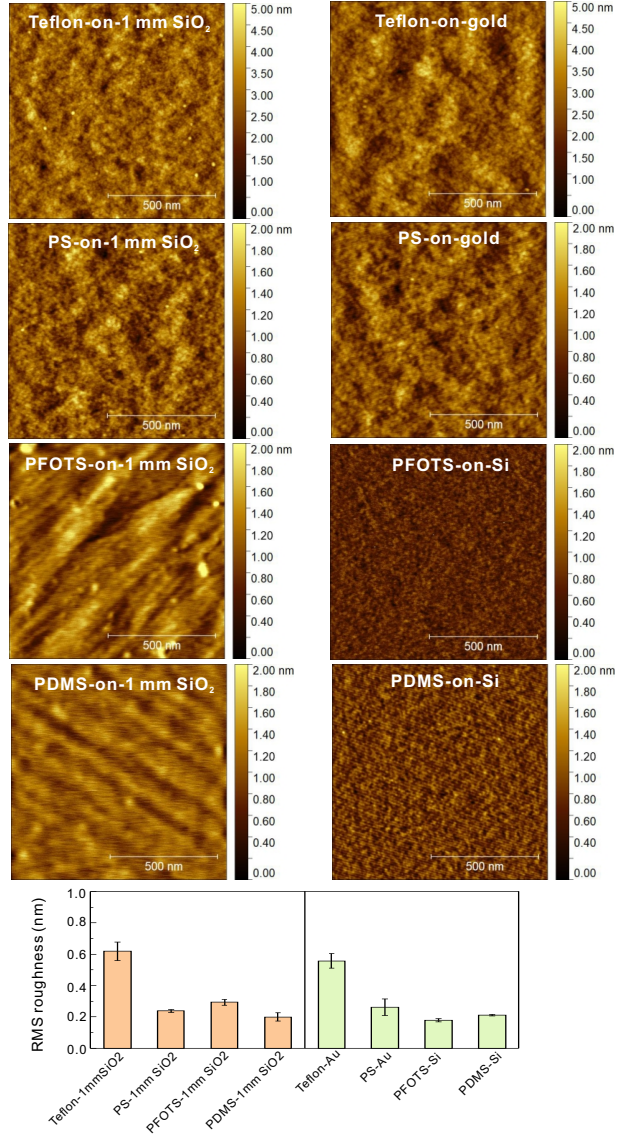

FIG. S1. Morphology and root-mean-square (RMS) roughness of all surfaces.

## S4: Change of solid-liquid interfacial tension by electrowetting.

When a drop slides on the Teflon-quartz surface with a back electrode, the dielectric layer between the drop and the electrode comprises a 50 nm thick Teflon coating and 1 mm thick quartz. The capacitance ($C$) is

$$\frac{1}{C} = \frac{1}{C_p} + \frac{1}{C_q} = \frac{d_p}{A\varepsilon_0\varepsilon_p} + \frac{d_q}{A\varepsilon_0\varepsilon_q} \approx \frac{d_q}{A\varepsilon_0\varepsilon_q}. \tag{S1}$$

Here, $C_p$ and $C_q$ are the capacitance due to the polymer coating and the quartz substrate, respectively. The relative dielectric permittivity of the quartz is $\varepsilon_q = 4.5$. A is the contact area of the drop of $A \approx 17\,\text{mm}^2$. Thus, $C \approx 0.68\,\text{pF}$. In addition, the electrostatic voltage $\Delta U$ is given by

$$\Delta U = \frac{Q}{C}, \tag{S2}$$

where $Q$ is the drop charge. In our previous measurements [S1], after 4 cm sliding on a Teflon-quartz surface with velocities of $0.2 - 0.5\,\text{m/s}$, the drop charge was $\approx 0.7\,\text{nC}$. From that, we obtain $\Delta U \approx 1.03\,\text{kV}$. According to equation

S2, the corresponding change of solid-liquid interfacial tension is $\Delta\gamma_{\mathrm{SL}} = -(C\Delta U^2)/2A \approx 10.6\,\mathrm{mN/m}$.

### S5: Influence of tungsten wire on the sliding drop measurement.

A tungsten wire (diameter$= 0.025\,\mathrm{mm}$, from Alfa Aesar, USA) was spanned parallel to surfaces with a height of $\approx 1\,\mathrm{mm}$. We first recorded a $30\,\mu\mathrm{L}$ $1\,\mathrm{mM}$ NaCl aqueous solution drop sliding on a $40°$ tilted Teflon-gold surface in contact with the tungsten wire. To study the influence of the grounded tungsten wire on the drop motion, we then lifted the wire above the drop and recorded another drop sliding along the same path again without contact with the wire (Fig. S2a). The velocities, dynamic advancing contact angles, and dynamic receding contact angles of the two drops are almost the same (Fig. S2b-d). Thus, we conclude that the influence of the tungsten wire itself on the sliding drop measurement can be ignored.

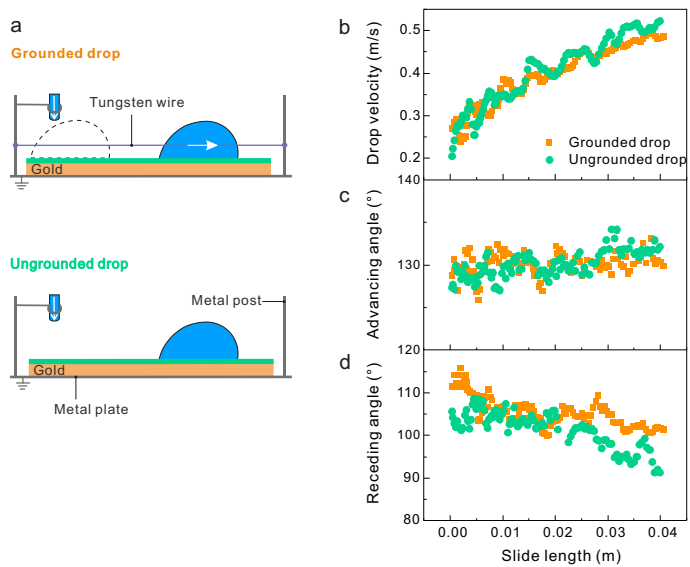

FIG. S2. Assessment of the influence of the grounded tungsten wire on the sliding drop measurement. (a) Schematics showing grounded and ungrounded drops on Teflon-gold surfaces. (b-d) Velocities, dynamic advancing contact angles, and dynamic receding contact angles versus slide length of grounded (orange) and ungrounded (green) drops sliding on the Teflon-gold surface.

### S6: Drop-number dependence for ungrounded drops.

We measured multiple successive ungrounded drops as well. On the Teflon-gold surfaces, the sliding drops were not influenced by slide electrification. Thus, there is no drop-number dependence of the drop velocity and the dynamic contact angles (Fig. S3), further indicating that the Teflon-gold surface is a good reference system for experiments without charge accumulation. On the Teflon-quartz surface, both the drop velocity and the dynamic contact angles depend on the drop number (Fig. S4). Compared with the reference, the dynamic advancing angle for the $1^{\mathrm{st}}$ ungrounded drop decreases from $125°$ to $105°$ after $4\,\mathrm{cm}$ sliding, $20°$ lower than the reference, which is within the expectation based on the electrowetting theory. The dynamic receding angle of the $1^{\mathrm{st}}$ ungrounded drop decreases from $93°$ to $67°$ after $4\,\mathrm{cm}$ sliding, $30°$ lower than the reference. The reduction of the dynamic contact angle at the rear side is more than at the front side for the $1^{\mathrm{st}}$ ungrounded drop, supporting the idea that deposited surface charges affect the contact angle besides electrowetting. For the $100^{\mathrm{th}}$ ungrounded drop, a reduction of the dynamic advancing and receding contact angles occurs as well, which, however, is less than the reduction for the $1^{\mathrm{st}}$ drop. This is consistent with less charge accumulation inside the drop and therefore a reduced influence of electrowetting. Interestingly, the reduction of the dynamic contact angle for the $1^{\mathrm{st}}$ ungrounded drop is even less than the one of the $1^{\mathrm{st}}$ grounded drop. Based on this, we speculate that preventing drop from charging facilitates charge deposition on solid surfaces.

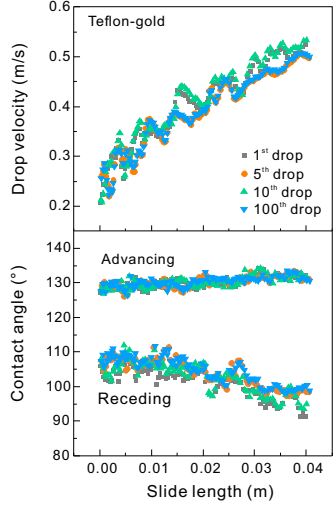

FIG. S3. Drop velocity, dynamic advancing angle, and dynamic receding contact angle versus slide length for the $1^{st}$, $5^{th}$, $10^{th}$, and $100^{th}$ ungrounded drop sliding on the Teflon-gold surface.

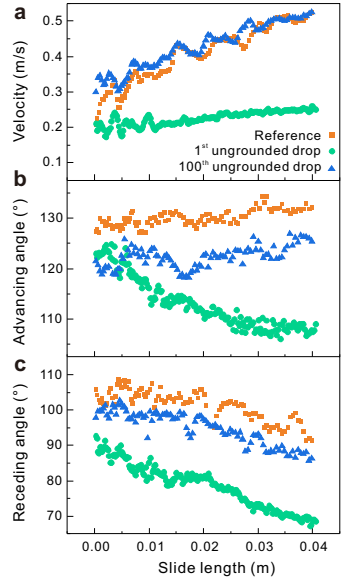

FIG. S4. Drop velocity, dynamic advancing angle, and dynamic receding contact angle versus slide length for the $1^{st}$ and $100^{th}$ ungrounded drop sliding on the Teflon-quartz surface.

## S7: Change of solid surface energy by surface charges.

Here, we derive a scaling relationship for the electrostatic correction to Young's equation due to the charges deposited at the rear end of a drop sliding along a surface. Two electrostatic effects contribute to the increase in free surface energy when charges are present: the self-energy of the individual charges, and the Coulomb interaction energy between the charges. The self-energy of a single ion of charge q at the interface between two dielectrics is $q^2/[4\pi\varepsilon_0(\varepsilon_r+1)a]$, where $a$ is the radius of the ion. Per unit area, this leads to an interfacial energy of $\Delta\gamma_S = q\sigma/[4\pi\varepsilon_0(\varepsilon_r+1)a]$. With a typical ionic radius of $a = 0.15\,\mathrm{nm}$ and measured charge densities of $\sigma = 10\,\mu\mathrm{C/m^2}$ [S1], the estimated increase in solid surface energy is only $17\,\mu\mathrm{N/m}$. This is too small to cause a substantial change in contact angle.

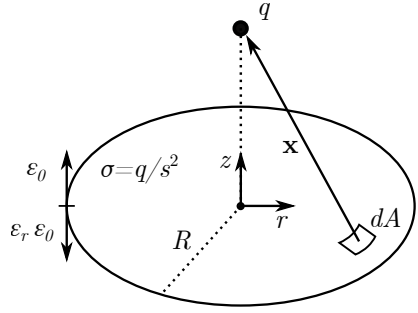

FIG. S5. Schematic representation of a charged circular area of radius $R$ and surface charge $\sigma = q/s^2$ at the interface of air and a dielectric used in the analytical calculations. A single charge $q$ is positioned centrally at a distance $z$. The distance from the single charge to an infinitesimal charged surface element $dA$ is indicated.

To quantify the Coulomb interaction energy of the charges on the surface, we analytically compute the work required to deposit an additional charge $q$ onto a surface carrying a charge density $d\sigma$. For an infinite charged plane, the electric field does not decay and the work becomes infinite. Consequently, the problem depends on the extension of the charged area. To this end, we consider a circular charged area of radius $R$ located at the interface of air and a dielectric with relative permittivity $\varepsilon_r$ (Fig. S5). The force $d\mathbf{F}$ between a single charge $q$ located at a distance $z$ above the center of the charged area and a differential surface element $dA$ of charge $d\sigma dA$ is

$$d\mathbf{F} = \frac{q d\sigma}{2\pi\varepsilon_0(1+\varepsilon_r)} \frac{\mathbf{x}}{|\mathbf{x}|^3} dA, \tag{S3}$$

where $\mathbf{x}$ is the distance between the surface element and the single charge. The total force on the single charge points in a wall-normal direction and is found by integrating $d\mathbf{F}$ over the charged area,

$$\mathbf{F} = \int_S d\mathbf{F} = \frac{q d\sigma}{2\pi\varepsilon_0(1+\varepsilon_r)} \int_S \frac{\mathbf{x}}{|\mathbf{x}|^3} dA. \tag{S4}$$

We introduce cylindrical coordinates originating at the center of the charged area and obtain for the normal force $F_n = \mathbf{F} \cdot \mathbf{e}_z$

$$F_n = \frac{q d\sigma}{2\pi\varepsilon_0(1+\varepsilon_r)} \int_0^{2\pi} \int_0^R \frac{zr}{(r^2+z^2)^{3/2}} dr d\phi = \frac{q d\sigma}{\varepsilon_0(1+\varepsilon_r)} \left(1 - \frac{1}{\sqrt{1+R^2/z^2}}\right). \tag{S5}$$

The work required to deposit the additional single charge on the surface is found by integrating $dW = -F_n dz$ from $z = \infty$,

$$-W_0 = \int_0^\infty dW = -\int_0^\infty F_n dz,$$
$$\longrightarrow W_0 = \frac{q d\sigma}{\varepsilon_0(1+\varepsilon_r)} \int_0^\infty \left(1 - \frac{1}{\sqrt{1+R^2/z^2}}\right) dz = \frac{q d\sigma R}{\varepsilon_0(1+\varepsilon_r)}. \tag{S6}$$

The total Coulomb interaction energy on the surface is given by the integral over the surface charge density from zero to $\sigma = q/s^2$, where s is the characteristic spacing of surface charges,

$$E_0 = \int_0^{q/s^2} \frac{qR}{\varepsilon_0(1+\varepsilon_r)} d\sigma = \frac{q\sigma R}{\varepsilon_0(1+\varepsilon_r)}. \tag{S7}$$

Finally, we divide $E_0$ by $s^2$ to get the area-specific surface energy due to Coulomb interaction

$$\frac{E_0}{s^2} = \Delta\gamma_S = \frac{\sigma^2 R}{\varepsilon_0(1+\varepsilon_r)}. \tag{S8}$$

Naturally, because the specific Coulomb interaction energy is $\propto \sigma^2$ and the specific self-energy is $\propto \sigma$, the Coulomb interaction energy dominates the increase in surface energy for $\sigma \gg q/(4\pi Ra) \approx 4\,\mathrm{nC/m^2}$.

**S8: Electrostatic length scale.**

In its simplest form, the electrostatic situation close to the receding contact line of a drop can be understood as an isopotential wedge next to a charged surface. The electrostatic potential distribution is governed by Laplace's equation, which does not contain an inherent length scale. Essentially, the observed change in contact angle is a consequence of Maxwell stresses deforming the liquid surface. However, from an equation without an intrinsic length scale, one would expect a deformation of the liquid surface that extends over the entire surface of the drop rather than a deformation occurring on such small scales that it only becomes visible as a change in contact angle. This raises the question about the mechanism that induces a sharp localization of the Maxwell stress in the close vicinity of the three-phase contact line.

This question is related to another problem, which is the singularity of the electric field strength at the tip of an isopotential wedge. It is well-known that the electric field strength diverges at this point, which raises the question about the physical mechanism that cuts off the singularity.

With respect to the latter question, different effects come into consideration. The idealized treatment of the liquid as a perfect conductor in the electrostatic problem implies that any Debye screening layers are regarded as infinitely thin. However, in an aqueous medium, counter charges are typically located in a diffuse layer with a thickness between 1 and 100 nm. Thus, treating the liquid surface as isopotential is only valid on length scales larger than the Debye length. This introduces an additional microscopic length scale that could resolve the singularity in the electric field.

A second effect that needs to be considered is dielectric breakdown. Humid air experiences dielectric breakdown at field strengths above about 2 MV/m. Electrostatic discharge thus constitutes a second mechanism that can screen the singularity in the electric field on the length scale where the breakdown field strength is exceeded. For relevant surface charges of the order of $\mu C/m^2$, both of these length scales are much smaller than the macroscopic length scale set by the drop radius and represent a scale that can no longer be resolved by optical imaging of a drop.

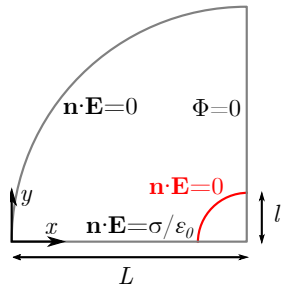

FIG. S6. Computational domain and boundary conditions. The horizontal boundary represents the charged solid surface and the vertical boundary the liquid surface. $L$ and $l$ are the macroscopic and microscopic length scales, respectively. The red circular boundary represents the implementation of the microscopic length scale.

To assess whether this microscopic length scale resolves the singularity in the Maxwell stress and localizes it microscopically in a narrow region around the contact line, we numerically solve the electrostatic Laplace equation,

$$\nabla^2 \Phi = 0, \tag{S9}$$

with the electric potential $\Phi$ and the electric field $\mathbf{E} = -\nabla\Phi$, in the computational domain shown in Fig. S6. It represents the gas phase around a charged surface next to an isopotential wedge, where we exemplarily set the contact angle to 90°. On the horizontal boundary, we specify the surface charge density , and on the vertical boundary we set the potential to zero, without loss of generality. The circular arc far from the contact line is assigned a homogeneous Neumann boundary condition, which corresponds to a vanishing normal component of the electric field. The radius $L$ of the wedge represents the macroscopic length scale as set by the droplet radius. We introduce a microscopic length scale by removing a small section of radius $l$ close to the contact line (red, Fig. S6) and compare the results to the situation without a microscopic length scale, $l \to 0$ (gray, Fig. S6). Processes within the excluded section, like electrostatic discharge, are beyond the validity of the governing equation. We use the finite element solver Comsol Multiphysics, version 6.0. We ensure grid independence of the results with a microscopic length scale by systematic refinement, monitoring the Maxwell stress at the coordinates $(L, l)$. Naturally, the case without a microscopic length scale cannot be grid independent arbitrarily close to the singularity.

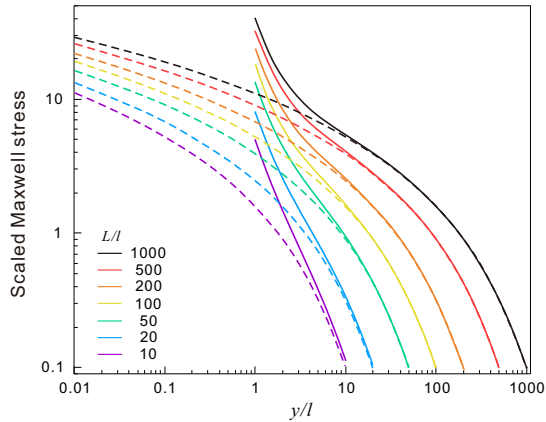

FIG. S7. Normal Maxwell stress scaled by $\sigma^2/\varepsilon_0$ along the liquid surface over the distance from the contact line $y/l$ for various macroscopic length scales $L/l$, indicated by different colors. Solid lines correspond to simulations with and dashed lines to simulations without the inclusion of the microscopic length scale. The microscopic length scale eliminates the singularity and localizes the Maxwell stress near the contact line.

In Fig. S7 we show the normal Maxwell stress on the liquid surface, scaled by $\sigma^2/\varepsilon_0$, over the distance from the contact line, $y/l$, for various ratios of the length scales $L/l$. Dashed lines indicate the reference case without a microscopic length scale and solid lines the case with a microscopic length scale. Without a microscopic length scale, the Maxwell stress becomes singular at the contact line. Introducing the microscopic length scale resolves the singularity and focuses the Maxwell stress around the point $y/l = 1$ to an interval up to about $y/l < 10$. Thus, the introduction of a scale $l$ has the effect that the forces deforming the liquid surface largely act in a region extending over a similar scale. This effect is roughly independent of the macroscopic length, as it is present across two orders of magnitude of $L/l$, which was the range considered in the simulations. We conclude that there is a microscopic length scale that eliminates the electrostatic singularity, for example the Debye length or the scale related to the onset of electrostatic discharge in air. The latter is of the order of $100\,\mathrm{nm}$ for relevant surface charge densities of $\approx 10\,\mathrm{\mu C/m^2}$. Due to the strong localization of the Maxwell stress on scales comparable to the microscopic length scale, the electrostatic forces result in a deformation of the liquid surface that is measurable as a contact angle change that can be modeled using Young's equation.

## S10. The influence of different polymer coatings.

The reduction of dynamic contact angles occurs for quartz substrates with different hydrophobic coatings. We measured aqueous drops containing $1\,\mathrm{mM}$ NaCl sliding on a PS-quartz surface, PDMS-quartz surface, and PFOTS-quartz surface with velocities of $0.2 - 0.3\,\mathrm{m/s}$. The corresponding values $\cos(\theta') - \cos(\theta)$ range between 0.05 and 0.25 (Fig. S8).

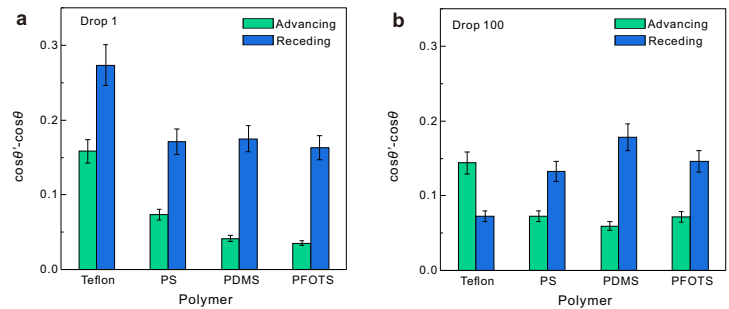

FIG. S8. $\cos(\theta') - \cos(\theta)$ for (a) the 1$^{\text{st}}$ and (b) the 100$^{\text{th}}$ ungrounded drop containing 1 mM NaCl with velocities of $0.2 - 0.3$ m/s on quartz substrates with different hydrophobic coatings. $\theta$ and $\theta'$ are the dynamic contact angles without and with the influence of slide electrification.

## S10: Influence of salt type, salt concentration, and drop volume.

We measured the dynamic advancing angles and the dynamic receding angles of ungrounded drops with different salts, different salt concentrations, and drop volumes on 40° tilted Teflon-gold surfaces. The velocity-dependent dynamic advancing and receding angles are similar for all drops (Fig. S10). Thus, we conclude that the influence of salt type, salt concentration, and drop volume on the dynamic contact angles for the Teflon-gold surfaces is ignorable.

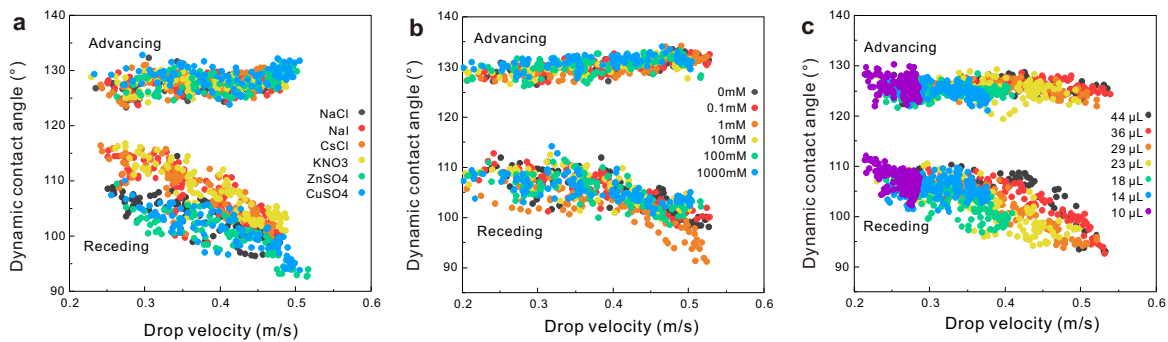

FIG. S9. Velocity-dependent dynamic contact angles for the 1$^{\text{st}}$ ungrounded drop for (a) different salts, (b) different NaCl concentrations, and (c) different drop volumes on 40° tilted Teflon-gold surfaces.

We measured ungrounded drops with different salts, NaCl concentrations, and volumes on the Teflon-quartz surface as well. For comparison, we calculated $\cos(\theta') - \cos(\theta)$ based on the Young-Lippmann equation at a velocity regime of $0.3 - 0.4$ m/s for the 1$^{\text{st}}$ ungrounded drop. Similar to grounded drops, both the dynamic advancing angle and the dynamic receding contact angle were reduced for drops with different salts (Fig. S10a). In addition, $\cos(\theta') - \cos(\theta)$ first increases and then decreases with the increase in salt concentration (Fig. S10b). $\cos(\theta') - \cos(\theta)$ stays almost constant as a function of drop volume for the dynamic advancing angle, while it increases for the dynamic receding angle (Fig. S10c). The dynamic advancing angle of the 1$^{\text{st}}$ ungrounded drop is influenced only by drop charging (electrowetting effect). Because of the compensation between contact area-dependent drop charge and contact area-dependent capacitance, the electrowetting effect is independent of drop volume. By contrast, on the receding side, both drop charging and surface charging affect the contact angle. The influence from surface charging is proportional to the drop radius, thus, $\cos(\theta') - \cos(\theta)$ increases with drop volume.

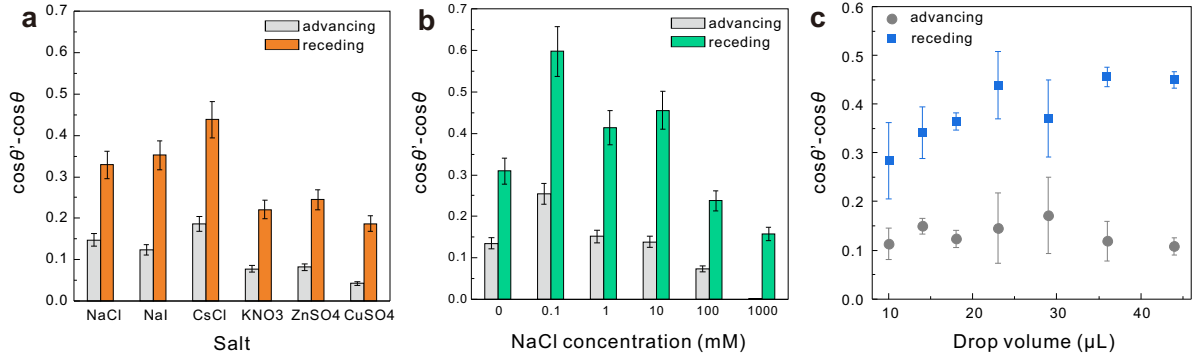

FIG. S10. $\cos(\theta') - \cos(\theta)$ for the 1$^{\text{st}}$ ungrounded drop with different salts (a), with different NaCl concentrations (b), and different volumes of 1 mM NaCl aqueous solution sliding on the Teflon-quartz surfaces with a velocity of $0.3 - 0.4 \, \text{m/s}$. $\theta$ and $\theta'$ are the dynamic contact angles without and with the influence of slide electrification.

---

* X.L. and A.D.R. contributed equally to this work.
† hardt@nmf.tu-darmstadt.de
‡ butt@mpip-mainz.mpg.de



\end{document}